\documentclass[floatfix,aps,pre,twocolumn,showpacs]{revtex4}
\usepackage{graphics}
\usepackage{eepic,epic}
\usepackage[pdftex]{graphicx}               

\begin{document}
\addtolength{\topmargin}{0.5in}

\title{Column collapse of granular rods}


\author{M.~Trepanier}
\author{Scott~V.~Franklin}
\email[]{svfsps@rit.edu}
\homepage[]{http://piggy.rit.edu/franklin/}
\affiliation{Department of Physics, Rochester Institute of Technology}

\date{\today}

\begin{abstract}
We investigate the collapse of granular rodpiles as a function of
particle (length/diameter) and pile (height/radius) aspect ratio.  We
find that, for all particle aspect ratios below 24, there exists a
critical height $H_l$ below which the pile never collapses,
maintaining its initial shape as a solid, and a second height $H_u$
above which the pile always collapses.  Intermediate heights between
$H_l$ and $H_u$ collapse with a probability that increases linearly
with increasing height.  The linear increase in probability is
independent of particle length, width, or aspect ratio.  When piles
collapse, the runoff scales as a piecewise power-law with pile height,
with $r_f\sim \tilde H^{1.2\pm 0.1}$ for pile heights below $\tilde H_c
\approx 0.74$ and $r_f\approx H^{0.6\pm0.1}$ for taller piles.

\end{abstract}

\maketitle
\section{introduction}

The collapse and flow of granular materials is fundamental to
catastrophic events such as avalanches, landslides, and pyroclastic
flows. Understanding how initially stationary piles of granular
material collapse, how far the material travels before stopping, and
the final pile geometry, is an important step in efforts to deal with
the practical consequences of such natural phenomena.

Previous experimental investigations in granular column collapse
\cite{PhysRevE_72_041301, Lube04} have focused on cylindrical piles
(initial radius $r_0$) of varying height $H$.  The radius of the
collapsed pile, $r_f$, scaled as $\tilde r \equiv (r_f -r_0)/r_0$,
shows a power-law dependence on dimensionless height $\tilde H \equiv
H/r_0$.  There is a crossover in the power-law exponent when $\tilde
H$ becomes larger than a critical height $\tilde H_c$, although there
is not consensus on the specific value of $\tilde H_c$.  Lube et
al. \cite{PhysRevE_72_041301, Lube04} found
\begin{equation}
\tilde r = \left \{ \begin{array}{c l}
\tilde H^{1.} & \tilde H < 1.7\\
\tilde H^{0.5} & \tilde H > 1.7 \end{array} \right .
\end{equation}
for a variety of irregular granular materials, independent of surface
roughness.  The crossover was also observed to coincide with a slight
change in the final pile shape.

Lajeunesse et al. \cite{lajeunesse:2371, lajeunesse:103302} studied
more regular materials (smooth glass beads) and also observed a change
in scaling of the runoff, but found the critical height to be $\tilde
H_c \approx 0.74$.  The collapse of low ($\tilde H < 0.74$) piles
resulted in a mesa-like final state, with some fraction of the top
surface remaining horizontal.  The horizontal plateau shrank with
increasing $\tilde H$, finally disappearing for $\tilde H > 0.74$.
Both Lube and Lajeunesse found little change in behavior for a variety
of particle types or surface roughness.  Their studies encompassed
fine and coarse quartz sand, salt, couscous, rice, sugar, and smooth
glass beads.

\begin{figure}[h]
\includegraphics[height=0.65in]{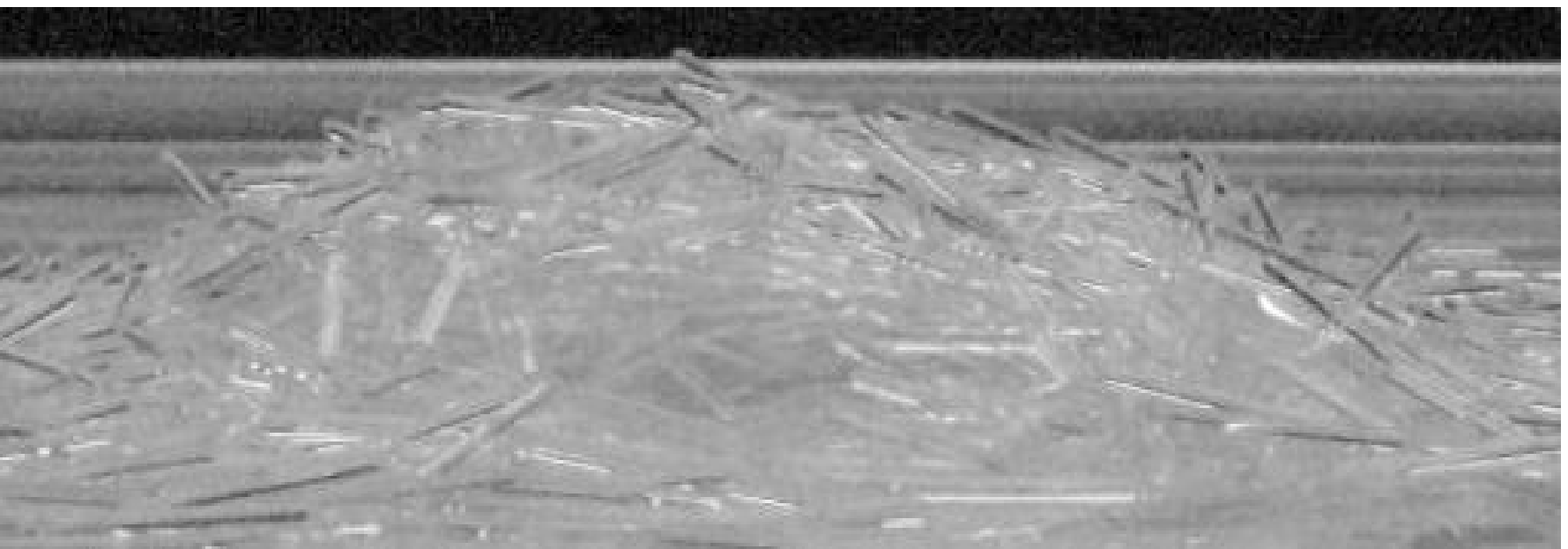}\hfill \includegraphics[height=.65in]{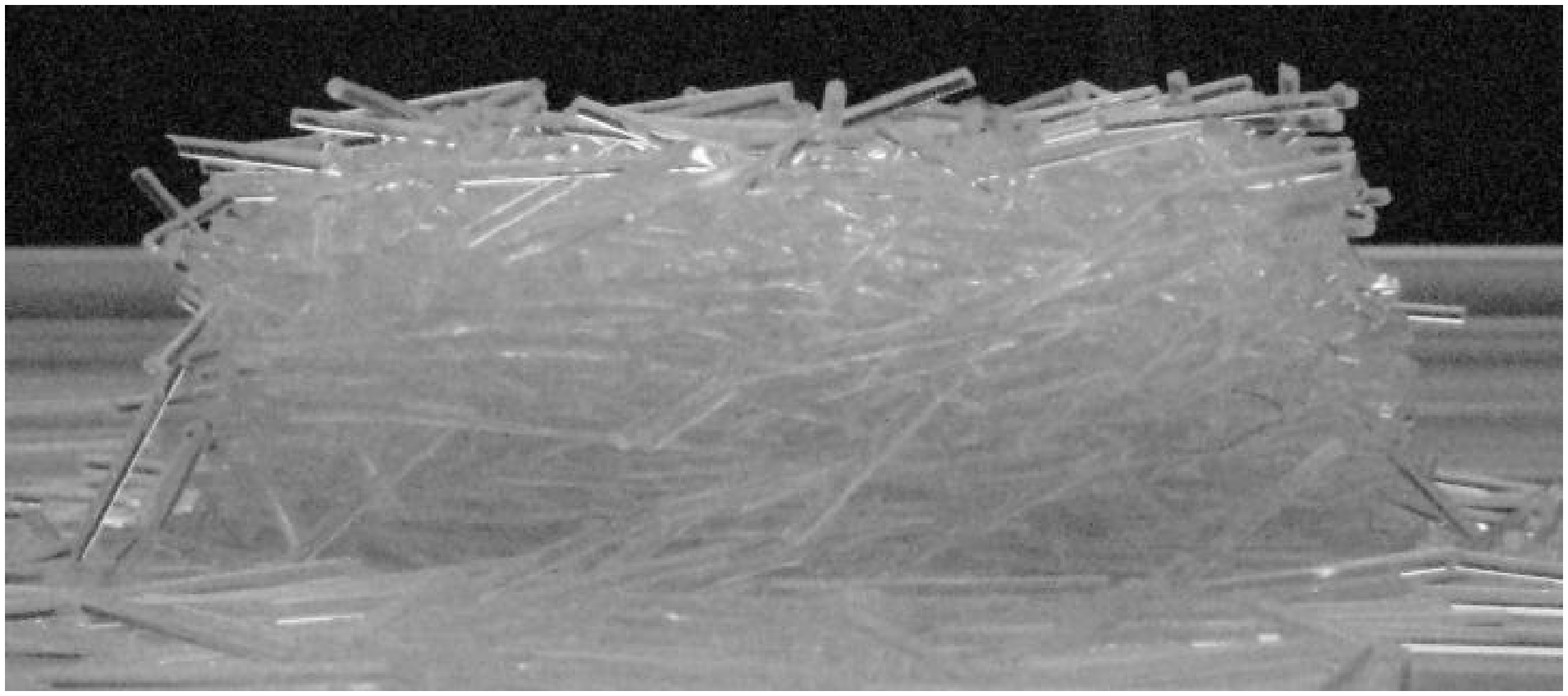}
\caption{\label{collapse}Rod-like piles of moderate height may
  collapse (left), but may also retain their original, cylindrical
  shape (right).  Shown are results from two experiments on aspect
  ratio $\alpha=16$ particles with identical initial heights.  We find
  the probability of collapse increases linearly with pile height.}
\end{figure}

It is well known that rod-like granular materials of sufficient aspect
ratio $\alpha \equiv L/d$ can support $90^\circ$ angles of repose, and
thus piles of these granular materials maintain the shape of the
container in which they were initially formed (c.f. figure
~\ref{collapse} (right)).  Philipse \cite{Philipse1, Philipse2,
  Philipse3, PhysRevE.67.051301} first studied this solid plug-like
behavior, and observed that particles with aspect ratio greater than
$\approx 35$ could not be poured from a container, but rather emerged
as a solid, a behavior has been assumed to be due to geometric
entanglement of the particles.

This has led to a great deal of research on rods of various aspect
ratio.  Villarruel et al. \cite{Villarruel} found that rods of aspect
ratio $\approx 4$ constrained to a cylinder would, when tapped, align
with the cylinder walls and compact.  Lumay and Vandewalle
\cite{Lumay04} extended this experiment to much larger aspect ratios,
and developed a 2d lattice model that reproduced the salient features
of the compaction curves.  Lumay and Vandewalle subsequently
\cite{Lumay06} used a two-dimensional experiment to highlight the
importance of rotation in rod rearrangements.  Blouwolff and Fraden
\cite{Blouwolff_Fraden_06} investigated the contact number in 3d
packings, and Stokely et al. \cite{Stokely} looked at static
properties of 2d packings.  Pournin et al. \cite{GranMatt_7_119_2005,
  ramaioli:021304} have written discrete element method (DEM)
simulations of spherocylindrical particles that show crystallization
and ordering under various excitations.  Recently, Hidalgo et
al. \cite{PhysRevLett.103.118001} looked at the role particle shape
plays on stress propagation through a granular packing, while Azema
and Radj\"\i \cite{PhysRevE.81.051304} looked at stress-strain
relations of rod packings under shear.  Finally, there has been some
attention paid to the nature of random packings.  Wouterse et
al. \cite{Wouterse07} looked at the microstructure of random packings
of spherocylinders and Zeravcic et al. \cite{Zeravcic09} have looked
at the excitations of random ellipsoid packings at the onsite of
jamming.  Despite all this attention, there have been no structural
characteristics identified that explain the increased rigidity of
large aspect-ratio particles.

Desmond et al. \cite{Desmond05} studied the transition from
classically granular to solid-like behavior by measuring the force on
a small object pulled through a rodpile.  They found
characteristically granular slip-stick behavior for low aspect ratio
particles, while large aspect ratio particles responded as a solid,
moving as a single coherent unit.  Desmond et al. were the first to
observe the existence of a transition region separating the granular
and solid behaviors.  Moderate aspect-ratio particles displayed both
granular and solid behavior, often alternating between the two.
Typically a pile would briefly display solid-like behavior and move
coherently, before collapsing around the intruder.  Desmond et
al. mapped the behavior in a two-dimensional parameter space of
particle aspect ratio and inverse container size, but there was no
quantitative measure of the transition region.

\section{Experimental Setup}
Our experimental apparatus follows that of Lube and Lajeunesse
\cite{PhysRevE_72_041301, Lube04, lajeunesse:2371, lajeunesse:103302}.
An acrylic cylinder with inner radius $r_0$ is filled with material to
height $H$ and then quickly lifted vertically.  Two cylinder radii
were tested, $r_0=5.4\pm0.1$~cm and $7.3\pm 0.1$~cm.  We observed no
dependence on cylinder radius and, unless otherwise noted, all
reported results are from the smaller cylinder.  The cylinder was
filled by dropping particles in from the top.  Two orthogonal
cross-beams spanned the top entrance to the cylinder and randomized
the particles' orientation as they fell into the cylinder.  The
cross-beams also broke up any clumps that existed prior to pouring.

The cylinder was raised with speeds ranging from 1-10 cm/s; little
dependence on cylinder speed was observed.  After the pile had
collapsed, the distance from the cylinder's central axis to the pile
edge, defined as the midpoint of the farthest particle in contact with
the main pile, was measured in four directions ($\pm x, \pm y$) and
averaged.

Rods were cut from acrylic and Teflon rod stock.  Particle diameters
were $D=0.16\pm 0.01 - 0.95 \pm$ cm, lengths $L=2.5\pm0.2 - 7.6\pm
0.2$ cm, resulting in a range of aspect ratio from $2.6-47.5$.  The
nominal coefficient of friction acrylic is $\mu_{acr}=0.8$; that of
Teflon $\mu_T=0.04$.  Ordinary sand was tested as a control, and
compared with the results of Lube and Lajeunesse et al.

\section{Scaling of Runoff with Height}

\begin{figure}[h]
\begin{center}
\includegraphics[angle=-90]{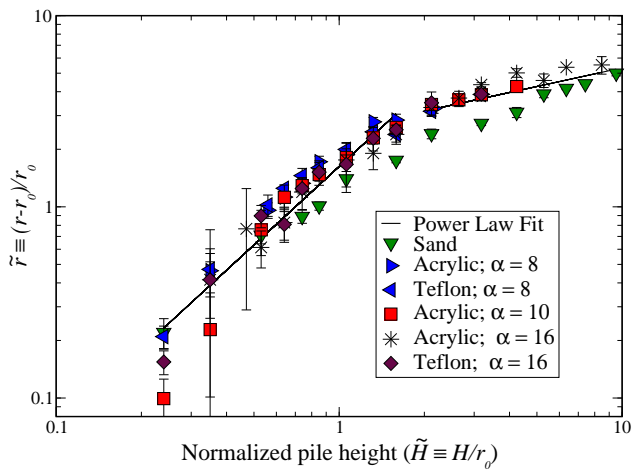}
\caption{\label{fig1}The runoff distance $\tilde r$ of a collapsed
  granular pile shows a power-law dependence on initial height $\tilde
  H$, with a transition between linear and square-root scaling.  Both
  radius and pile height $\tilde H$ are made dimensionless by the
  initial pile radius $r_0$.  For $\tilde H < \approx 1.1$ the runoff
  increases linearly with $\tilde H$, whereas for larger piles the
  runoff increases as $\tilde H^{1/2}$ (solid lines are power law
  fits).  Both scaling behaviors, as well as the transition point,
  arise from conservation of volume arguments that assume all material
  outside an internal, stable cone collapses.}
\end{center}
\end{figure}

Following Lube, we define the runoff as the change in pile radius,
normalized by the initial radius: $\tilde r \equiv (r_f-r_0)/r_0.$ As
shown in Fig.~\ref{fig1}, sand, acrylic rods of aspect ratio 8, 10,
and 16, and Teflon rods of aspect ratio 8 and 16 all show a power-law
dependence on normalized pile height $\tilde H \equiv H/r_0$ with
change in exponent at a critical height $\tilde H_c$.  To determine
the the exponents and crossover height we perform a 3-parameter
least-squares fit of the data to the piece-wise power-law function 
\begin{equation}\label{eq1}
\tilde r = \left \{ \begin{array}{c l}
\tilde H^{\alpha_1} & \tilde H < \tilde H_c\\
\tilde H^{\alpha_2} & \tilde H > \tilde H_c \end{array} \right .
\end{equation}
The minimization is done analytically for $\alpha_1$ and $\alpha_2$
and numerically for $\tilde H_c$.  We have done this fitting both for
every individual data run, averaging the resulting exponents and
transition point, as well as for the aggregate collection of all data.
Both analyses fall within the same standard-deviation uncertainty for
the parameters, with $\alpha_1=1.2\pm0.1, \alpha_2=0.6\pm0.1$, and
$\tilde H_c=1.1 \pm 0.3$, or, putting these values into Eq.~\ref{eq1}
\begin{equation}
\tilde r = \left \{ \begin{array}{c l}
\tilde H^{1.2\pm0.1} & \tilde H < 1.1\pm0.3\\
\tilde H^{0.6\pm0.1} & \tilde H > 1.1\pm0.3 \end{array} \right .
\end{equation}

The range of heights tested is only about a decade for the lower
region, and less than a decade for the upper region, and so we only
assert that our results are consistent with the findings of Lube et
al.~\cite{Lube04} and Lajeunesse et al.~\cite{lajeunesse:2371,
  lajeunesse:103302}.

As worked out by Lajeunesse \cite{lajeunesse:2371}, all of the
behavior in Fig.~\ref{fig1} --- the two power-law scaling exponents
and the crossover between the two --- can be explained by conservation
of volume, assuming the material does not compact as it collapses.
The crossover height occurs when the dimensionless pile height is such that
$$\tilde H_0=\tan \theta_c,$$ where $\theta_c$ is the dynamic angle of
repose.  Our observed transition height of $\tilde H_0 = 1.1 \pm 0.3$
implies an angle of repose of about $\theta_c \approx 47^\circ \pm
7^\circ$, slightly higher than that found by Lajeunesse for round and
irregular particles, indicating a slightly increased structural
stability of rodpiles.

To conclude this section, the runoff of rods of even moderate aspect
ratios follows similar scaling as that of sand.  The comparable
transition point indicates a similar angle of repose suggesting that,
when the pile collapses, the cylindrical particles roll much like
spheres.  The square-root scaling of runoff at large pile heights
indicates that the dynamics of pile collapse are also similar, with an
immobile inner core setting the final pile height.

\section{Solidity of moderate aspect-ratio particles}

One of the more striking behaviors of large aspect-ratio granular
material is their ability to maintain the shape of their initial
container.  As expected, piles formed from particles with aspect ratio
($L/d$) greater than 24 never collapsed, maintaining their cylindrical
shape.  While 24 is slightly lower than the value commonly assumed
necessary to maintain plug-like behavior (\cite{Philipse1,
  Philipse2,Blouwolff_Fraden_06, Desmond05}), it is consistent with
the finding of Desmond et al. that finite-size container effects can
induce plug-like behavior at lower aspect ratios \cite{Desmond05}.
Surprisingly, however, we find that this solid-like behavior manifests
in moderate aspect-ratio particles as well (see Fig.~\ref{collapse}
(right)).

The behavior of piles formed from particles with intermediate aspect
ratios $4<\alpha<24$ could not be predicted a priori, with experiments
with visually equivalent initial conditions producing dramatically
different results.  Shown in Fig.~\ref{collapse} are two piles that
resulted with particles of aspect ratio $\alpha=16$ with relatively
low initial pile heights.  As can be seen, one pile collapsed,
resulting in a mound formation with a relatively well-defined angle of
repose, while the other did not, remaining cylindrical.  When the pile
collapsed, its runoff obeyed the scaling behavior shown in
Fig.~\ref{fig1}.

\begin{figure}[h]
\begin{center}
 \includegraphics[angle=-90]{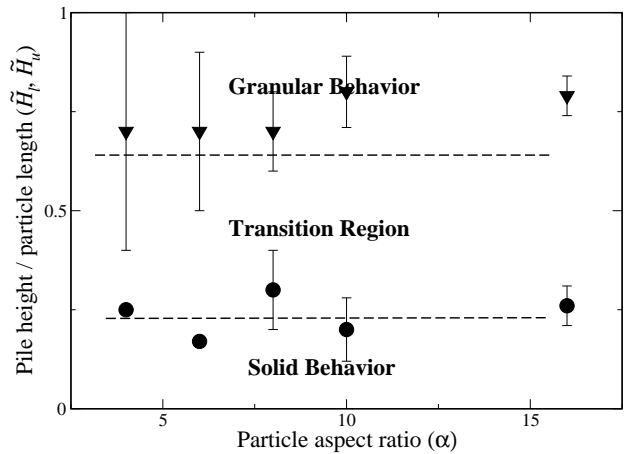}
\caption{\label{trans}The critical heights that determine whether a
  pile acts as a solid or granular material, scaled by particle
  length, are independent of aspect ratio.  The solid and granular
  regions are separated by a transition region in which the pile has a
  probability of collapsing, although there are no visible differences
  between piles that collapse and those that do not.}
\end{center}
\end{figure}

For all aspect ratios below 24, a critical height exists above which
the pile always collapsed.  Additionally, for each of these aspect
ratios there exists a minimum height below which the pile never
collapsed.  Increasing the particle aspect ratio increased these
critical heights, and so the initial radius does not seem the
appropriate quantity by which to characterize this behavior.  We find,
however, that these critical heights are the same for all aspect
ratios tested when scaled by the particle length $L$.  This is shown
in Fig.~\ref{trans} for acrylic particles.  The dashed lines exist to
guide the eye and reveal qualitative changes in behavior when the pile
height is less than $L/4$, between $L/4$ and $3L/4$, and greater than
$3L/4$.  It should be noted that the data in Fig.~\ref{trans} come
from a variety of particle diameters and initial cylinder radii, and
so argue that it is the particle length that determines the maximum
pile height.  Reducing particle friction narrowed, but did not
eliminate, the transition region.

\begin{figure}[h]
\begin{center}
\includegraphics[angle=-90]{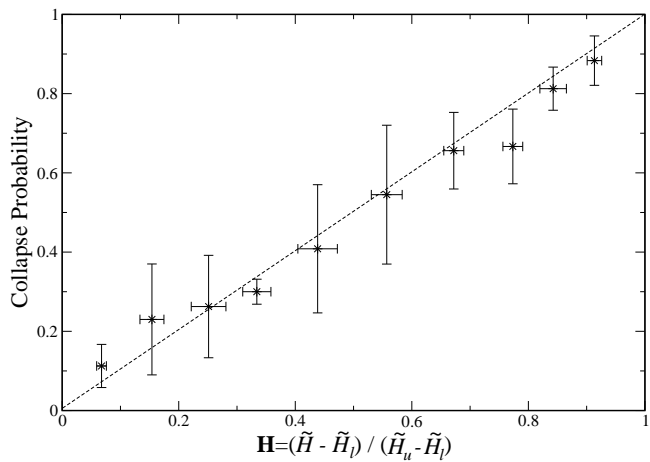}
\caption{\label{figlast}The probability of collapse within the
  transition region increases linearly with pile height scaled by the
  critical heights.  This behavior is independent of particle length,
  width, and aspect ratio.  The normalization by critical heights is
  done so that the origin is the lower transition point and (1,1) is
  the higher transition point.} 
\end{center}
\end{figure}

The probability for a pile to collapse increases as one moves through
the transition region.  We use the upper and lower critical heights
$\tilde H_l$ and $\tilde H_u$ to parametrize the height, defining an
height order parameter that is 0 at $\tilde H_l$ and 1 at $\tilde
H_u$:
\begin{eqnarray}\label{hscal}{\rm \bf H}\equiv \frac{\tilde H - \tilde H_l}{\tilde H_u - \tilde
  H_l}.
\end{eqnarray} In Figure~\ref{figlast} we plot the collapse probability as a
function of this order parameter {\bf H} for all of our data,
encompassing the complete range of aspect ratios that collapse (4-20)
and two different particle diameters.  Figure~\ref{figlast} shows
that, when scaled as per Eq.~\ref{hscal}, all the data collapse onto a
straight line between (0,0) (the point at which no piles collapse) to
(1,1) (where all piles collapse).
The data in Fig.~\ref{figlast} are grouped by height into bins of
scaled width 0.1, corresponding to the horizontal error bars.  The
vertical error bars represent one standard deviation of the average
collapse probability of all data in that bin.  

Figures \ref{trans} and \ref{figlast} represent the first quantitative
description of the transition region in which the material can behave
either as a solid or as a granular material.  This region is unique to
large aspect-ratio materials, and appears to exist even when the
aspect ratio is moderate ($\sim 4$).  Computational work is underway
to determine what internal characteristics (e.g. distribution of
forces or percolating force chains) might be responsible for the
different behavior in superficially similar piles.

The region of bistability was observed in two different cylinder
radii, and the critical heights are normalized by the cylinder
radius.  Both cylinders were larger than a particle length; it is not
clear how piles formed in extremely small containers (compared to the
particle length) would respond.  In this case particles could not
orient horizontally, and this may reduce the stability of any pile formed.
Desmond et al. \cite{Desmond05} found that extremely large containers
allowed for a more granular response to an intruder, but it is
not clear that the same behavior exists in this geometry.  We saw no
decrease in pile rigidity in the larger cylinder tested.

\section{Conclusions}
We have investigated the collapse of granular columns consisting of
cylindrical rods.  Interestingly, when the pile collapses, it does so
in a manner very similar to that of spheres or irregular sand. The
runoff distance, when scaled by the initial pile radius, initially
scales linearly with the normalized pile height.  This corresponds to
the partial collapse of the pile, involving only the material at the
outer edges.  The final pile shape is that of a truncated cone, with
outer radius determined by the angle of repose of the material.  At a
critical pile height, also determined by the angle of repose, there is
a crossover to a different scaling, where the runoff grows as the
square-root of scaled height.  This indicates an immobile inner cone
of material, at the angle of repose, and that the final pile height is
set by the initial radius.  Both the scaling behavior, and the
location of the cross-over height, are quantitatively consistent with
earlier work on spheres, and so we conclude that, when rodpiles
collapse, they are quantitatively the same as ordinary sand or
spheres.

Large-aspect ratio particles do display a fundamentally different
behavior from that of spheres, however, in their ability to maintain
the shape of their initial container, essentially acting as a solid.
We have shown that the onset of this behavior with increasing aspect
ratio is not sudden, but manifests in even moderate aspect ratios
under suitable experimental geometries.  For a given aspect ratio
there exist critical pile heights that demarcate the solid and
granular behaviors, and the transition region separating the two.  The
transition region is characterized by a probability of solid/granular
response, with a linear transition between the two.  The transition
heights, when scaled by particle length, are independent of particle
aspect ratio.  This behavior persists until the aspect ratio is quite
large ($>24$), at which all piles are essentially solid.  Future
research will investigate the internal characteristics that determine
the pile response, as well as the behavior of the critical heights
with increasing particle aspect ratio and pile radius.

\begin{acknowledgments}
This research was supported in part by the National Science Foundation
under Award No. DMR-0706353 and in part by an award from the Research
Corporation; Scott Franklin is a Cottrell Scholar of Research
Corporation.  Acknowledgment is also made to the Donors of the
American Chemical Society Petroleum Research Fund for support of this
research.
\end{acknowledgments}


\bibliography{references}

\begin{thebibliography}{20}
\expandafter\ifx\csname natexlab\endcsname\relax\def\natexlab#1{#1}\fi
\expandafter\ifx\csname bibnamefont\endcsname\relax
  \def\bibnamefont#1{#1}\fi
\expandafter\ifx\csname bibfnamefont\endcsname\relax
  \def\bibfnamefont#1{#1}\fi
\expandafter\ifx\csname citenamefont\endcsname\relax
  \def\citenamefont#1{#1}\fi
\expandafter\ifx\csname url\endcsname\relax
  \def\url#1{\texttt{#1}}\fi
\expandafter\ifx\csname urlprefix\endcsname\relax\def\urlprefix{URL }\fi
\providecommand{\bibinfo}[2]{#2}
\providecommand{\eprint}[2][]{\url{#2}}

\bibitem[{\citenamefont{Lube et~al.}(2005)\citenamefont{Lube, Huppert, Sparks,
  and Freundt}}]{PhysRevE_72_041301}
\bibinfo{author}{\bibfnamefont{G.}~\bibnamefont{Lube}},
  \bibinfo{author}{\bibfnamefont{H.~E.} \bibnamefont{Huppert}},
  \bibinfo{author}{\bibfnamefont{R.~S.~J.} \bibnamefont{Sparks}},
  \bibnamefont{and} \bibinfo{author}{\bibfnamefont{A.}~\bibnamefont{Freundt}},
  \bibinfo{journal}{Physical Review E} \textbf{\bibinfo{volume}{72}},
  \bibinfo{pages}{041301} (\bibinfo{year}{2005}).

\bibitem[{\citenamefont{Lube et~al.}(2004)\citenamefont{Lube, ad~R.~Stephen
  J.~Sparks, and Hallworth}}]{Lube04}
\bibinfo{author}{\bibfnamefont{G.}~\bibnamefont{Lube}},
  \bibinfo{author}{\bibfnamefont{H.~E.~H.} \bibnamefont{ad~R.~Stephen
  J.~Sparks}}, \bibnamefont{and} \bibinfo{author}{\bibfnamefont{M.~A.}
  \bibnamefont{Hallworth}}, \bibinfo{journal}{Journal of Fluid Mechanics}
  \textbf{\bibinfo{volume}{508}}, \bibinfo{pages}{175} (\bibinfo{year}{2004}).

\bibitem[{\citenamefont{Lajeunesse et~al.}(2004)\citenamefont{Lajeunesse,
  Mangeney-Castelnau, and Vilotte}}]{lajeunesse:2371}
\bibinfo{author}{\bibfnamefont{E.}~\bibnamefont{Lajeunesse}},
  \bibinfo{author}{\bibfnamefont{A.}~\bibnamefont{Mangeney-Castelnau}},
  \bibnamefont{and} \bibinfo{author}{\bibfnamefont{J.~P.}
  \bibnamefont{Vilotte}}, \bibinfo{journal}{Physics of Fluids}
  \textbf{\bibinfo{volume}{16}}, \bibinfo{pages}{2371} (\bibinfo{year}{2004}),
  \urlprefix\url{http://link.aip.org/link/?PHF/16/2371/1}.

\bibitem[{\citenamefont{Lajeunesse et~al.}(2005)\citenamefont{Lajeunesse,
  Monnier, and Homsy}}]{lajeunesse:103302}
\bibinfo{author}{\bibfnamefont{E.}~\bibnamefont{Lajeunesse}},
  \bibinfo{author}{\bibfnamefont{J.~B.} \bibnamefont{Monnier}},
  \bibnamefont{and} \bibinfo{author}{\bibfnamefont{G.~M.} \bibnamefont{Homsy}},
  \bibinfo{journal}{Physics of Fluids} \textbf{\bibinfo{volume}{17}},
  \bibinfo{eid}{103302} (pages~\bibinfo{numpages}{15}) (\bibinfo{year}{2005}),
  \urlprefix\url{http://link.aip.org/link/?PHF/17/103302/1}.

\bibitem[{\citenamefont{Philipse}(1996)}]{Philipse1}
\bibinfo{author}{\bibfnamefont{A.~P.} \bibnamefont{Philipse}},
  \bibinfo{journal}{Langmuir} \textbf{\bibinfo{volume}{12}},
  \bibinfo{pages}{1127} (\bibinfo{year}{1996}).

\bibitem[{\citenamefont{Philipse and Verberkmoes}(1997)}]{Philipse2}
\bibinfo{author}{\bibfnamefont{A.~P.} \bibnamefont{Philipse}} \bibnamefont{and}
  \bibinfo{author}{\bibfnamefont{A.}~\bibnamefont{Verberkmoes}},
  \bibinfo{journal}{Physica A} \textbf{\bibinfo{volume}{235}},
  \bibinfo{pages}{186} (\bibinfo{year}{1997}).

\bibitem[{\citenamefont{Philipse and Kluijtmans}(1997)}]{Philipse3}
\bibinfo{author}{\bibfnamefont{A.~P.} \bibnamefont{Philipse}} \bibnamefont{and}
  \bibinfo{author}{\bibfnamefont{S.~G. J.~M.} \bibnamefont{Kluijtmans}},
  \bibinfo{journal}{Physica A} \textbf{\bibinfo{volume}{274}},
  \bibinfo{pages}{516} (\bibinfo{year}{1997}).

\bibitem[{\citenamefont{Williams and Philipse}(2003)}]{PhysRevE.67.051301}
\bibinfo{author}{\bibfnamefont{S.~R.} \bibnamefont{Williams}} \bibnamefont{and}
  \bibinfo{author}{\bibfnamefont{A.~P.} \bibnamefont{Philipse}},
  \bibinfo{journal}{Phys. Rev. E} \textbf{\bibinfo{volume}{67}},
  \bibinfo{pages}{051301} (\bibinfo{year}{2003}).

\bibitem[{\citenamefont{Villarruel et~al.}(2000)\citenamefont{Villarruel,
  Lauderdale, Mueth, and Jaeger}}]{Villarruel}
\bibinfo{author}{\bibfnamefont{F.~X.} \bibnamefont{Villarruel}},
  \bibinfo{author}{\bibfnamefont{B.~E.} \bibnamefont{Lauderdale}},
  \bibinfo{author}{\bibfnamefont{D.~M.} \bibnamefont{Mueth}}, \bibnamefont{and}
  \bibinfo{author}{\bibfnamefont{H.~M.} \bibnamefont{Jaeger}},
  \bibinfo{journal}{Physical Review E} \textbf{\bibinfo{volume}{61}},
  \bibinfo{pages}{6914} (\bibinfo{year}{2000}).

\bibitem[{\citenamefont{Lumay and Vandewalle}(2004)}]{Lumay04}
\bibinfo{author}{\bibfnamefont{G.}~\bibnamefont{Lumay}} \bibnamefont{and}
  \bibinfo{author}{\bibfnamefont{N.}~\bibnamefont{Vandewalle}},
  \bibinfo{journal}{Physical Review E} \textbf{\bibinfo{volume}{70}},
  \bibinfo{pages}{051314} (\bibinfo{year}{2004}).

\bibitem[{\citenamefont{Lumay and Vandewalle}(2006)}]{Lumay06}
\bibinfo{author}{\bibfnamefont{G.}~\bibnamefont{Lumay}} \bibnamefont{and}
  \bibinfo{author}{\bibfnamefont{N.}~\bibnamefont{Vandewalle}},
  \bibinfo{journal}{Physical Review E} \textbf{\bibinfo{volume}{74}},
  \bibinfo{pages}{021301} (\bibinfo{year}{2006}).

\bibitem[{\citenamefont{Blouwolff and Fraden}(2006)}]{Blouwolff_Fraden_06}
\bibinfo{author}{\bibfnamefont{J.}~\bibnamefont{Blouwolff}} \bibnamefont{and}
  \bibinfo{author}{\bibfnamefont{S.}~\bibnamefont{Fraden}},
  \bibinfo{journal}{Europhysics Letters} \textbf{\bibinfo{volume}{76}},
  \bibinfo{pages}{1095} (\bibinfo{year}{2006}).

\bibitem[{\citenamefont{Stokely et~al.}(2003)\citenamefont{Stokely, Diacou, and
  Franklin}}]{Stokely}
\bibinfo{author}{\bibfnamefont{K.}~\bibnamefont{Stokely}},
  \bibinfo{author}{\bibfnamefont{A.}~\bibnamefont{Diacou}}, \bibnamefont{and}
  \bibinfo{author}{\bibfnamefont{S.~V.} \bibnamefont{Franklin}},
  \bibinfo{journal}{Physical Review E} \textbf{\bibinfo{volume}{67}},
  \bibinfo{pages}{051302} (\bibinfo{year}{2003}).

\bibitem[{\citenamefont{Pournin et~al.}(2005)\citenamefont{Pournin, WEber,
  Tsukahara, Ferrez, Ramaioli, and Libling}}]{GranMatt_7_119_2005}
\bibinfo{author}{\bibfnamefont{L.}~\bibnamefont{Pournin}},
  \bibinfo{author}{\bibfnamefont{M.}~\bibnamefont{WEber}},
  \bibinfo{author}{\bibfnamefont{M.}~\bibnamefont{Tsukahara}},
  \bibinfo{author}{\bibfnamefont{J.~A.} \bibnamefont{Ferrez}},
  \bibinfo{author}{\bibfnamefont{M.}~\bibnamefont{Ramaioli}}, \bibnamefont{and}
  \bibinfo{author}{\bibfnamefont{T.~M.} \bibnamefont{Libling}},
  \bibinfo{journal}{Granular Matter} \textbf{\bibinfo{volume}{7}},
  \bibinfo{pages}{119} (\bibinfo{year}{2005}).

\bibitem[{\citenamefont{Ramaioli et~al.}(2007)\citenamefont{Ramaioli, Pournin,
  and Liebling}}]{ramaioli:021304}
\bibinfo{author}{\bibfnamefont{M.}~\bibnamefont{Ramaioli}},
  \bibinfo{author}{\bibfnamefont{L.}~\bibnamefont{Pournin}}, \bibnamefont{and}
  \bibinfo{author}{\bibfnamefont{T.~M.} \bibnamefont{Liebling}},
  \bibinfo{journal}{Physical Review E (Statistical, Nonlinear, and Soft Matter
  Physics)} \textbf{\bibinfo{volume}{76}}, \bibinfo{pages}{021304}
  (\bibinfo{year}{2007}).

\bibitem[{\citenamefont{Hidalgo et~al.}(2009)\citenamefont{Hidalgo, Zuriguel,
  Maza, and Pagonabarraga}}]{PhysRevLett.103.118001}
\bibinfo{author}{\bibfnamefont{R.~C.} \bibnamefont{Hidalgo}},
  \bibinfo{author}{\bibfnamefont{I.}~\bibnamefont{Zuriguel}},
  \bibinfo{author}{\bibfnamefont{D.}~\bibnamefont{Maza}}, \bibnamefont{and}
  \bibinfo{author}{\bibfnamefont{I.}~\bibnamefont{Pagonabarraga}},
  \bibinfo{journal}{Phys. Rev. Lett.} \textbf{\bibinfo{volume}{103}},
  \bibinfo{pages}{118001} (\bibinfo{year}{2009}).

\bibitem[{\citenamefont{Az\'ema and Radja\"\i{}}(2010)}]{PhysRevE.81.051304}
\bibinfo{author}{\bibfnamefont{E.}~\bibnamefont{Az\'ema}} \bibnamefont{and}
  \bibinfo{author}{\bibfnamefont{F.}~\bibnamefont{Radja\"\i{}}},
  \bibinfo{journal}{Phys. Rev. E} \textbf{\bibinfo{volume}{81}},
  \bibinfo{pages}{051304} (\bibinfo{year}{2010}).

\bibitem[{\citenamefont{Wouterse et~al.}(2007)\citenamefont{Wouterse, Williams,
  and Philipse}}]{Wouterse07}
\bibinfo{author}{\bibfnamefont{A.}~\bibnamefont{Wouterse}},
  \bibinfo{author}{\bibfnamefont{S.~R.} \bibnamefont{Williams}},
  \bibnamefont{and} \bibinfo{author}{\bibfnamefont{A.~P.}
  \bibnamefont{Philipse}}, \bibinfo{journal}{Journal of Physics: Condensed
  Matter} \textbf{\bibinfo{volume}{19}}, \bibinfo{pages}{406215}
  (\bibinfo{year}{2007}),
  \urlprefix\url{http://stacks.iop.org/0953-8984/19/i=40/a=406215}.

\bibitem[{\citenamefont{Zeravcic et~al.}(2009)\citenamefont{Zeravcic, Xu, Liu,
  Nagel, and van Saarloos}}]{Zeravcic09}
\bibinfo{author}{\bibfnamefont{Z.}~\bibnamefont{Zeravcic}},
  \bibinfo{author}{\bibfnamefont{N.}~\bibnamefont{Xu}},
  \bibinfo{author}{\bibfnamefont{A.~J.} \bibnamefont{Liu}},
  \bibinfo{author}{\bibfnamefont{S.~R.} \bibnamefont{Nagel}}, \bibnamefont{and}
  \bibinfo{author}{\bibfnamefont{W.}~\bibnamefont{van Saarloos}},
  \bibinfo{journal}{EPL (Europhysics Letters)} \textbf{\bibinfo{volume}{87}},
  \bibinfo{pages}{26001} (\bibinfo{year}{2009}),
  \urlprefix\url{http://stacks.iop.org/0295-5075/87/i=2/a=26001}.

\bibitem[{\citenamefont{Desmond and Franklin}(2006)}]{Desmond05}
\bibinfo{author}{\bibfnamefont{K.}~\bibnamefont{Desmond}} \bibnamefont{and}
  \bibinfo{author}{\bibfnamefont{S.~V.} \bibnamefont{Franklin}},
  \bibinfo{journal}{Physical Review E} \textbf{\bibinfo{volume}{73}},
  \bibinfo{pages}{031306} (\bibinfo{year}{2006}).

\end{thebibliography}
\end{document}